\documentclass[12pt]{article}

\usepackage{scicite}

\usepackage{times}

\usepackage{graphicx}
\usepackage[utf8]{inputenc}
\usepackage{amsmath}
\usepackage{verbatim}
\usepackage{amssymb}
\usepackage{listings}
\usepackage{color}
\usepackage{amsmath}
\usepackage{mathtools}
\usepackage{color}
\usepackage{chemformula}
\usepackage{siunitx}
\usepackage{xspace}
\usepackage{pifont}
\usepackage[inline]{enumitem}
\newcommand{\cmark}{\ding{51}}
\newcommand{\xmark}{\ding{53}}
\DeclareSIUnit{\angstrom}{\AA}
\DeclareSIUnit{\calorie}{cal}
\newcommand\subfig[2]{{Fig.~\ref{#1}{#2}}}
\newcommand\subcap[1]{{(#1):}}

\newcommand{\fig}[1]{Fig.~\ref{#1}}

\newcommand{\quot}[1]{``#1''}

\newcommand{\OCAL}{\mathcal{O}}  
\newcommand{\expb}[1]{\exp \glb #1 \grb} 
\newcommand{\loga}[2][]{\log^{#1}\! \gla #2 \gra}  

\newcommand{\gla}{\,}  
\newcommand{\gra}{}  
\newcommand{\glb}{\left(}  
\newcommand{\grb}{\right)}  

\newcommand\bigObs[1]{\ensuremath{\OCAL  (#1)}}

\newcommand{\JF}{\textsc{JeLLyFysh}\xspace}
\newcommand{\LA}{\textsc{Lammps}\xspace}
\renewcommand\subcap[1]{{(#1)}}

\makeatletter
\let\NAT@parse\undefined
\makeatother
\usepackage{hyperref}

\newenvironment{sciabstract}{%
	\begin{quote} \bf}
	{\end{quote}}

\title{Molecular simulation from modern statistics: Continuous-time,
continuous-space, exact}

\author
{Philipp H\"ollmer,$^{1}$ A.~C.~Maggs,$^{2}$ Werner Krauth$^{3\ast}$\\
	\\
	\normalsize{$^{1}$Physikalisches Institut and Bethe Center for Theoretical 
	Physics,}\\
	\normalsize{University of Bonn, Nussallee 12, 53115 Bonn, Germany}\\
	\normalsize{$^{2}$CNRS UMR7083, ESPCI Paris, Université PSL,}\\ 
	\normalsize{10 Rue Vauquelin, 75005 Paris, France} \\
	\normalsize{$^{3}$Laboratoire de Physique de l’Ecole normale 
        supérieure, ENS, Université PSL,}\\
	\normalsize{CNRS, Sorbonne Université, Université de Paris-Cité, Paris, 
France 
	}\\
	\\
	\normalsize{$^\ast$To whom correspondence should be addressed; E-mail:  
	werner.krauth@ens.fr.}
}

\date{}

\begin{document} 
	
	
	\baselineskip24pt

	\maketitle 
	
	
\begin{sciabstract}
In a world made of atoms, the computer simulation of molecular systems, such
as proteins in water, plays an enormous role in science. Software packages
that perform these computations have been developed for decades. In molecular
simulation, Newton's equations of motion are discretized and long-range
potentials are treated through cutoffs or spacial discretization, which
all introduce approximations and artifacts that must be controlled
algorithmically. Here, we
introduce a paradigm for molecular simulation that is based on modern concepts
in statistics and is rigorously free of discretizations, approximations,
and cutoffs. Our demonstration software reaches a break-even point
with traditional molecular simulation at high precision.
We stress the
promise of our paradigm as a gold standard for critical applications and as a
future competitive approach to molecular simulation.
\end{sciabstract}
	
\paragraph*{One-sentence summary:} A rigorous paradigm for exact non-reversible 
Markov processes is benchmarked for classical long-range-interacting water.\\

\noindent
The fact that all matter consists of atoms has been described as the greatest
insight of science~\cite{FeynmanVol1}. The consequence that matter can be
modeled on a computer by following the motion of its atoms leads to the founding
paradigm of molecular simulation. It tracks the dynamics and explores the
thermodynamic equilibrium of complex molecular systems, for example, a
peptide in an explicit water solution with tens of thousands of atoms,
all interacting through classical empirical potentials~\cite{Shaw2010}.
Molecular simulation is of enormous importance
to numerous fields ranging from biology and physics to 
engineering~\cite{SchlickBook,KarplusNobel}. Powerful
computer packages have been developed over
decades~\cite{Amber_2,CHARMM,GROMACS_1,LAMMPS,NAMD}. They
compute the forces on all atoms at discretized time steps and
then update the atomic positions and velocities to integrate
the classical equations of motion of molecular dynamics.
A voluminous literature is dedicated to the
analysis and control of time-discretization errors [see, e.g.,
Ref.~\cite{Hammonds2020}]. Thermostats, understood as
\quot{necessary evils}~\cite{Wong2016}, mimic the effect of a coupled thermal
reservoir and, in a symptomatic but non-curative treatment, hide the
accumulated errors. The limiting factor in molecular dynamics is the
computation of forces. The Lennard-Jones interaction is
typically cut off beyond a certain distance so that only a few neighbors exert a
force on any given atom.
The long-range nature of the
Coulomb potential, which must be preserved, is usually treated through fast
mesh-based Ewald 
methods~\cite{Hockney1988,Darden1993,EssmannPedersen1995,Petersen1995,Kohnke2020}
that solve the Poisson equation in
discretized space. Thermostats and cutoffs as well as the discretizations
inherent in long-range interactions
introduce approximations and
artifacts~\cite{Wong2016,Wennberg2013} that complex algorithms
aim to keep under control.

We present here an alternative paradigm for molecular simulation that is based
on modern concepts in statistics. It is rigorously exact from the start and, by
construction, strictly simulates the canonical ensemble without
thermostats. Straight-line trajectories of atoms in continuous Monte-Carlo time
are interrupted by \emph{events}. This non-reversible piecewise-deterministic
Markov process~\cite{Davis1984} violates the detailed-balance condition normally
associated with thermal equilibrium, but still samples the Boltzmann
distribution. Its use has led to spectacular speedups of local Markov chains in
statistical physics~\cite{Bernard2011}. Short- and long-range
potentials are handled without any cutoffs or discretizations and, as we
show in this paper, with competitive efficiency. The Boltzmann
weight $\pi=\exp(-\beta U)$ (with $\beta$ the inverse temperature and $U$ the
potential) is expressed as a factorized product $\pi=\prod_M
\exp(-\beta U_M)$ of statistically independent factors $M$ with factor
potentials $U_M$ with $\sum_M U_M = U$ that each depend only on a small
subensemble of atoms~\cite{Michel2014JCP}. Every factor stochastically generates
a time when the piecewise-deterministic motion must be interrupted. The minimum
of these times triggers an event, and determines the initial conditions for the
next piece. The total potential $U$ and the corresponding forces never need be
evaluated, yet the stationary state is rigorously the Boltzmann distribution.

We implement the event-driven paradigm in demonstration software for $N$
flexible SPC/Fw water molecules~\cite{WuTepperVoth2006} that interact with the
long-range Coulomb potential. As a benchmark, we concentrate on the electric
polarization~\cite{deLeeuw1980-1} (in other words, the total electric dipole
moment) that measures the capacity of individual water molecules to rotate and
determines the dielectric properties of water. We find that a particular
non-reversible Markov process greatly reduces autocorrelation times and
overcomes the slow diffusive dynamics of reversible Monte Carlo algorithms,
while the factorization does not penalize the dynamics in comparison to
molecular dynamics. The polarization decorrelates in a computer time that
scales as $N \loga{N}$, similar to mesh-based Ewald methods in molecular 
dynamics~\cite{Darden1993} but without their diverging precision-dependent 
prefactor. Our code reaches a break-even point with respect to a standard
molecular-dynamics code much below machine precision, and we point out
its great potential for improvement.

\paragraph*{Modern-statistics paradigm for SPC/Fw water}\mbox{}\\
In a molecular system with long-range interactions, the force on an atom depends
on the position of all other atoms, rendering its evaluation tedious unless one
introduces cutoffs or discretizes space. In contrast, we implement a
piecewise-deterministic Markov process through the event-chain Monte Carlo
algorithm, where a single atom moves at any given moment
~\cite{Michel2014JCP,Bernard2009,Krauth2021eventchain}. The deterministic motion
of this atom is interrupted by an event that stops it and sets off a similar
motion of a new atom. Factors $M$ in the SPC/Fw water model describe \ch{O-H}
bonds, the bending of \ch{H-O-H} opening angles, \ch{O-O} intermolecular
Lennard-Jones interactions, and the Coulomb interaction between two water
molecules. Each factor proposes an independent \emph{candidate event time}. The
minimum over all the candidate event times then realizes the
next event, and motion is transferred to another atom contributing to
$U_M$. This succession of  events, in which rejections are replaced by
transfers, distinguishes our formulation from the usual
Metropolis Monte Carlo algorithm~\cite{Metropolis1953}.

\begin{figure}[b]
	\centering
	\includegraphics{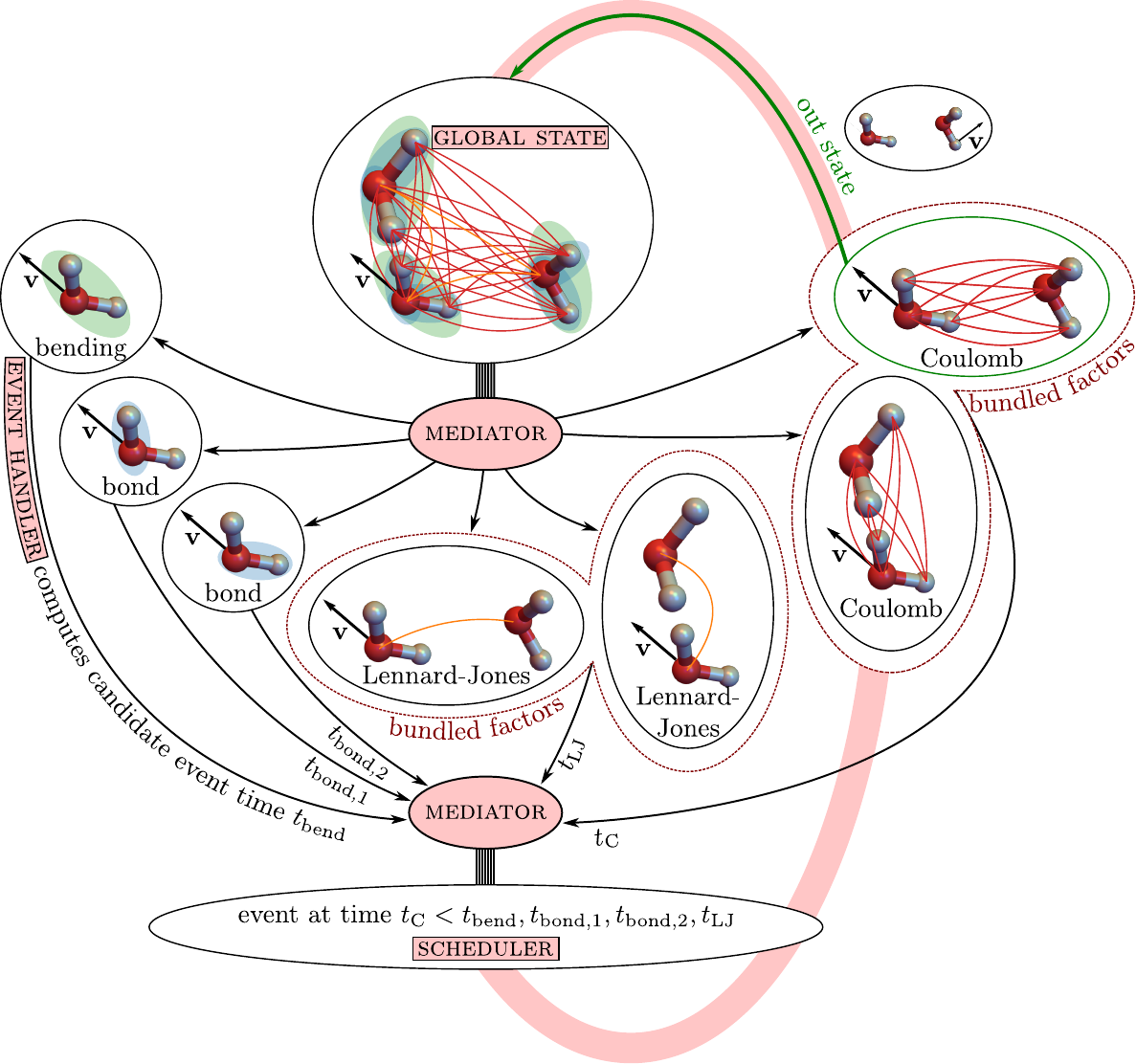}
	\caption{
	\textbf{\JF implementation of our Markov-chain paradigm.}
	The \emph{mediator}
    splits the \emph{global state} into
	statistically independent factors. Factors
    communicate	independent candidate event times, the earliest of which
    defines the next event. Factors for long-range interactions are bundled,
    so that the number of \emph{event handlers} remains limited.
    Candidate event times are collected by the mediator and then treated in
    the \emph{scheduler}. The factor triggering the event then updates the
    global state, again \emph{via} the mediator.}
	\label{fig:Mediator}
\end{figure}

Positions and velocities of atoms define the \emph{global state} of the physical
system. To impose coherency of the physical system, the global state is accessed
only through a central \emph{mediator}~\cite{GammaDesignPatterns1994} that
dispatches physically independent computations of candidate events to
\emph{event handlers}. A \emph{scheduler} weeds through candidate events and
identifies the unique event that provides the subsequent transfer of motion,
leading to an update of the global state (see~\fig{fig:Mediator}). The event
handlers within  the mediator architecture mirror the statistical independence
of the factors composing the physical system. This allows us to compose complex
interactions in a transparent and independent manner. A number of inequivalent
options have been constructed for the update of active particles within
event-chain Monte Carlo~\cite{Harland2017,Faulkner2018}. Similar flexibility is
possible in the updating of
velocities~\cite{Bernard2009,Klement2019,Michel2020,Hoellmer2022Dipoles}, as
well as in parallelizing~\cite{Hoellmer2020,Li2020} it. In this paper, we
replace the original straight variant by the substantially more efficient
Newtonian event-chain Monte Carlo~\cite{Klement2019,Hoellmer2022Dipoles}. It
requires no fine-tuning and again exactly samples the canonical Boltzmann
distribution. Modern statistics offers an even wider choice of options for the
management of events, the choice of factors, and the piecewise-deterministic
trajectories, that may well apply to molecular simulation in the
approximation-free non-reversible Markov-chain framework.

\begin{figure}[b]
\centering
\includegraphics[width=\textwidth]{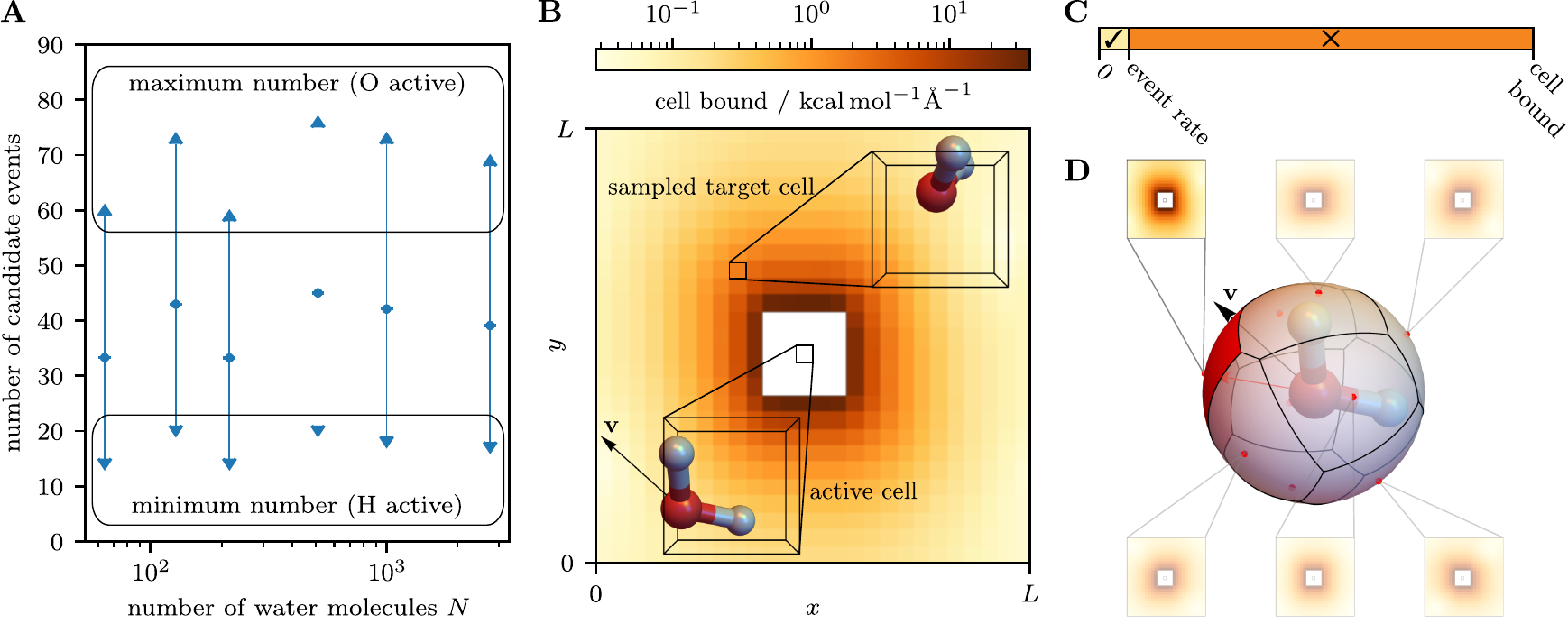}
\caption{\textbf{Long-range interactions with constant computer time per event.}
\subcap{\textbf{A}} The number of candidate events (event handlers)
is constant for increasing system sizes.
\subcap{\textbf{B}} Walker table from which a 
target cell is sampled according to its cell bound with respect to the active 
cell containing the moving atom.
\subcap{\textbf{C}} Different Walker tables for Fibonacci vectors on the unit
sphere. The active atom obtains cell bounds from the nearest vector.
\subcap{\textbf{D}} The thinning procedure confirms (\cmark) or rejects 
(\xmark) the event using the actual event rate of the molecules in the active 
and target cell.
}
\label{fig:CV}
\end{figure}

\paragraph*{Implementation for SPC/Fw water: Cell-veto---Fibonacci sphere} 
\mbox{}\\
In the event-driven implementation of our method (see \fig{fig:Mediator}), a
single atom moves among the other atoms and molecules, so that $\bigObs{N}$
factors are changing with time and in principle yield independent events that
would require sorting and managing. Modern statistics allows one to bundle most
of these factors and, in the SPC/Fw water model, the mediator interacts with
only
$\sim 50$ event handlers that propose  candidate event times to the scheduler
(see \subfig{fig:CV}{A}). The bundling allows the processing of each
event in
constant computer time (for large $N$) while treating the long-range
interactions without approximations.

In the cell-veto algorithm~\cite{KapferKrauth2016}, which we use for the
bundling, event rates for pairs of molecules interacting with the Coulomb
potential are upper-bounded by precomputed, time-independent bounds for these
molecules somewhere within a pair of cells (see \subfig{fig:CV}{B}). The full
set of these \emph{cell bounds} corresponds to the set of bundled factors of the
long-range interaction. Walker's method of
aliases~\cite{Walker1977AnEfficientMethod} conserves the cell bounds in a
\emph{Walker table}. We build separate such tables for multiple directions of
the velocity of the moving atom corresponding to \emph{Fibonacci vectors} on the
unit sphere (see \subfig{fig:CV}{C}). During the simulation, the actual velocity
of the moving atom is mapped to the closest Fibonacci vector. The set of cell
bounds in the corresponding table provides a single candidate event time for the
entire set of bundled factors, and Walker's method samples an associated single
cell bound---and thus a single associated factor---with constant algorithmic
complexity. The overestimation of the event rate by the cell bound is
corrected in a procedure akin to the thinning of non-homogeneous Poisson
processes~\cite{LewisShedler1979} by confirming the transfer of motion in the
event (see \subfig{fig:CV}{D}). This thinning is performed with the actual
positions of the atoms, leading to an exact treatment of the long-range
interaction that is independent of the set of cell bounds and the discretization
of space.

\begin{figure}[b]
\centering
\includegraphics[width=\textwidth]{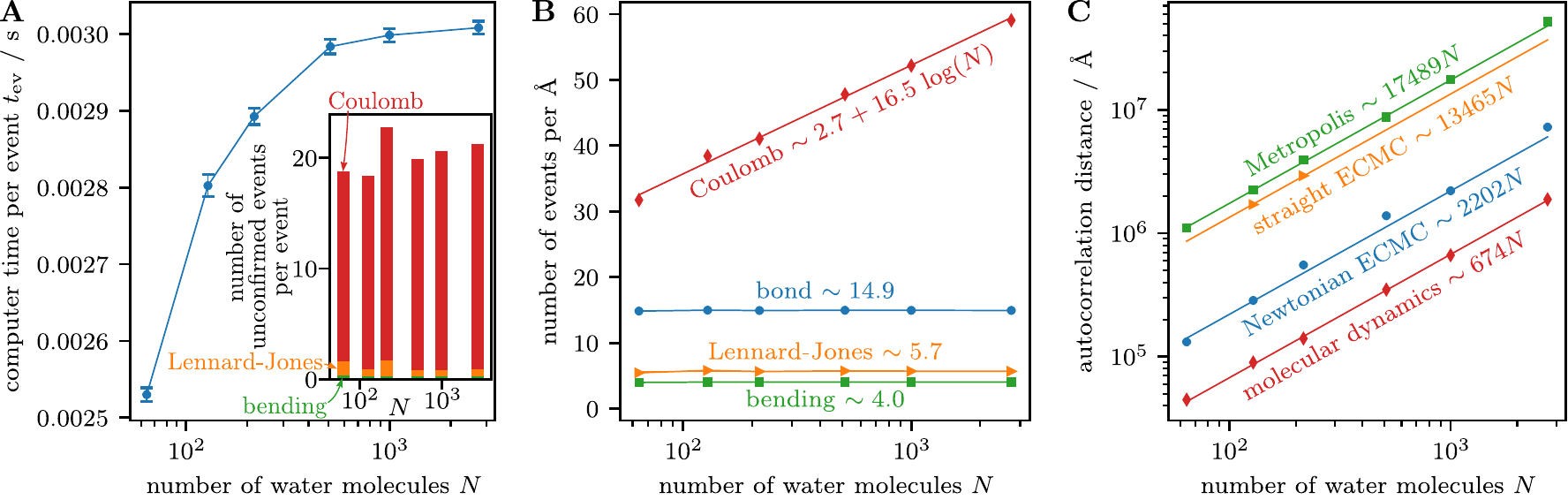}
\caption{\textbf{Event rates and decorrelation in the SPC/Fw water model.}
\subcap{\textbf{A}} Computer time per event in \JF.
Inset: Number of unconfirmed events per event for different factor
types.
\subcap{\textbf{B}} Event rate in \JF for different factor types.
\subcap{\textbf{C}} Distance to decorrelate the polarization
for different sampling
algorithms (for molecular dynamics: sum over the average displacements of
all atoms per time step).}
\label{fig:autocorrelation_distance}
\end{figure}

\paragraph*{Benchmarking for SPC/Fw water}\mbox{}\\
For our benchmark, we use \JF~\cite{Hoellmer2020} to sample configurations of 
SPC/Fw water mole\-cules
in a periodic box at standard density and temperature. We implement long-range
molecular Coulomb factors with Walker tables that we also adopt for the
Lennard-Jones interaction. We find that Newtonian event-chain Monte Carlo, for
large $N$, requires a computer time per event that remains constant (see
\subfig{fig:autocorrelation_distance}{A}). A large number of  unconfirmed
events stems from the overestimated cell bounds which, e.g., do not account for
the relative orientation of molecules. While the computer time per event is
constant,
the number of events per \AA ngstr\"om (that is, per unit Monte-Carlo time) 
increases
logarithmically for the Coulomb factors (see
\subfig{fig:autocorrelation_distance}{B}), as predicted by
theory~\cite{Faulkner2018}. In summary, our approach requires a computer time
scaling as $N \loga{N}$ to advance $N$ water molecules by a constant
distance. This matches the complexity of mesh-based Ewald
methods~\cite{Darden1993}, but without their slowdown as the precision is 
increased.

For concreteness, we compare the decorrelation of the polarization within \JF to
the \LA software on a single processor with default parameters and a $
\SI{1}{\femto \second}$ time step. To decorrelate this local quantity, both \LA
and \JF must move the atoms of any water molecule by a characteristic distance,
providing a first measure of efficiency
(see \subfig{fig:autocorrelation_distance}{C}). Different variants of  our
method vary in their efficiency, and the recently developed  Newtonian
event-chain Monte Carlo is an order of magnitude  faster than the straight
variant. The reversible Metropolis algorithm with single-atom moves [that also 
reaches an $N\log N$ 
scaling~\cite{Saunders2021} by using a recent variant of the fast multipole 
method~\cite{greenhard1987}], as implemented in the
\textsc{DL\_MONTE}
software package~\cite{DLMONTE}, is clearly inferior to our
non-reversible methods.

\begin{figure}[b]
\centering
\includegraphics{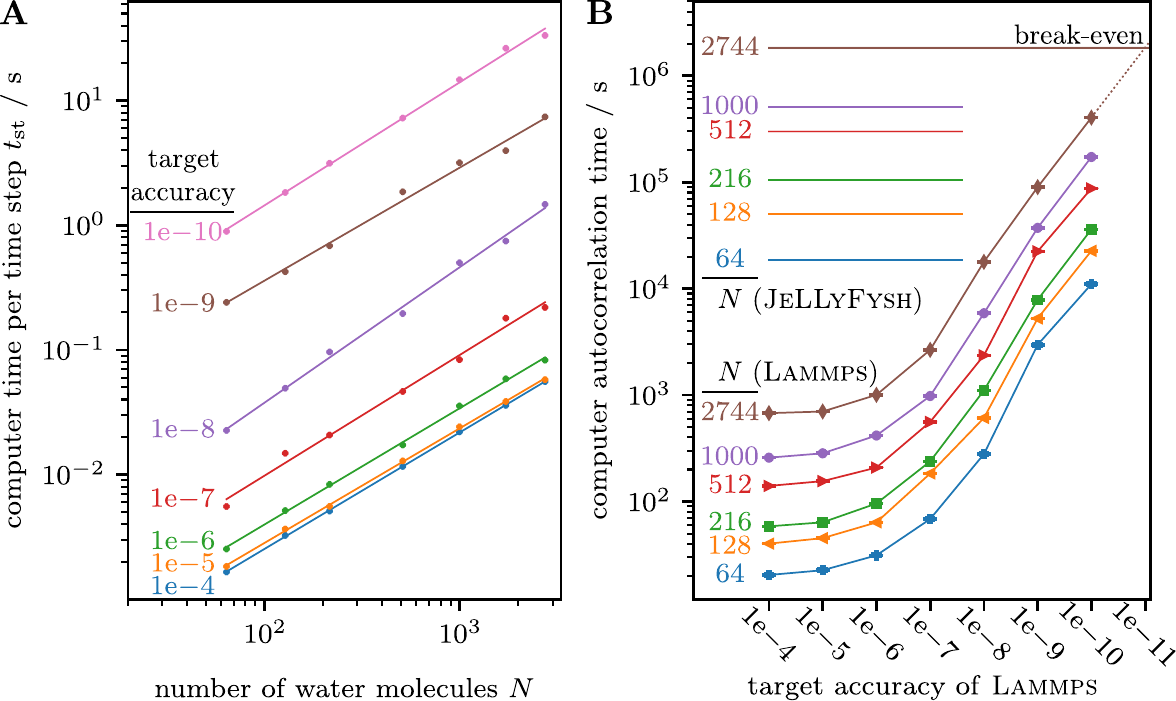}
\caption{\textbf{\LA--\JF benchmark for the SPC/Fw water model.} 
\subcap{\textbf{A}}
Computer time per step of \LA for different
target accuracies of its particle--particle particle--mesh solver.
\subcap{\textbf{B}} Computer time used by
\LA to decorrelate the polarization depends on the
target accuracy and the number of water molecules $N$. \JF
is exact up to machine precision. The break-even precision is
indicated.}
\label{fig:LammpsPrecisionPolarization}
\end{figure}

The mesh-based Ewald method implemented in \LA comes with a target accuracy that
is based on analytic error estimates obtained from a specific charge
distribution~\cite{Kolafa1992,Petersen1995,Deserno1998}. The charges of atoms
(which live in continuous space) are mapped onto the grid using an interpolation
scheme. Finer grids and higher interpolation orders yield higher target
accuracies. We estimate the necessary computer times by changing the grid
spacing,
using the \LA defaults (see \subfig{fig:LammpsPrecisionPolarization}{A}). \JF
uses the two-atom Coulomb interaction calculated with the historic Ewald
summation in continuous space. We tune it to machine precision without any
assumptions on the global charge distribution. \JF reaches
the break-even point to \LA well below machine precision (see
\subfig{fig:LammpsPrecisionPolarization}{B}), although it has only the status
of demonstration software.

\paragraph*{Discussion}\mbox{}\\
In this paper, we benchmarked an implementation of a modern-statistics paradigm
for molecular simulations in the standard SPC/Fw water model. The time
dependence of the corresponding Markov process differs from the physical
dynamics yet it exactly approaches thermal equilibrium on time scales that are
potentially faster than in nature. Its remarkable efficiency (that we expressed
as a $N \loga N$ computer time to decorrelate a local observable, the
polarization) is rooted in three paradoxes. First, the Markov process is
non-reversible (that is, effectively out-of-equilibrium), yet its steady state
coincides with the equilibrium Boltzmann distribution. In contrast to standard
Monte Carlo algorithms that satisfy the detailed-balance condition and only move
diffusively, it features finite probability flows, making it capable of moving
ballistically. This has already led to considerable speedups in a variety of
fields ranging from physics to statistics and machine learning [see, e.g.,
Refs~\cite{Krauth2021eventchain,Vanetti2018,Fearnhead2018,Bierkens2020}]. It
remains to be seen whether non-local observables, for example large-scale
hydrodynamic modes, macroscopic conformations, and order parameters, can
similarly benefit from non-reversibility in chemical physics. The second paradox
is that in our approach, the Boltzmann distribution $\expb{-\beta U}$ is sampled
without any approximation yet with great efficiency because  the potential $U$
(or its derivatives, the forces) are never evaluated. This sidesteps all the
problems with limited-precision calculations of energies and forces. The third
paradox is the bundling of $\mathcal{O}(N)$ independent decisions to interrupt
the straight-line trajectory into an expression that can be evaluated in
constant time. The Walker tables channel long-range Coulomb factors into a
single candidate event, which allows us to handle a complex decision (a
conjunction of $\mathcal{O}(N)$
factor-wise decisions of independent factor potentials $U_M$)
in a few operations, even in the $N \to \infty $ limit.

Our demonstration software is openly available and fully functional. As
discussed, it becomes competitive with a traditional molecular-dynamics code at
high intrinsic precision. Our method is exact from the very beginning, and
future research will be able to concentrate on the most efficient ones among a
large choice of cell bounds, factorizations, Fibonacci
vectors and variants of the piecewise-deterministic Markov processes.
Clearly, more interdisciplinary research from statistics to computational
chemistry will clarify whether this provides a sufficiently strong basis for an
alternative approach for practical molecular simulation.

With its guarantee for the unbiased sampling of the Boltzmann distributions, our
paradigm may serve as a gold standard for molecular simulation in general,
capable of identifying artifacts and approximations that may not have been
totally eliminated through the symptomatic algorithmic approach to molecular
dynamics. Furthermore, given the greater algorithmic freedom for the
Markov-chain approach than for molecular dynamics, it may actually become faster
than molecular dynamics. Several orders of magnitude in algorithmic
speed can certainly be gained by re-engineering our software, which would then
be able to tackle the peptide-in-water benchmark problem~\cite{Shaw2010} that
has had a major influence over the last decade. The great simplicity of our
approach and its present implementation in the \JF software may well facilitate
further developments.



\begin{thebibliography}{10}

\bibitem{FeynmanVol1}
R.~Feynman, R.~Leighton, M.~Sands, {\it The Feynman Lectures on Physics, Vol.
  I: Mainly Mechanics, Radiation, and Heat\/}, vol.~1 of {\it The Feynman
  Lectures on Physics\/} (Addison-Wesley Publishing Company, 1963).

\bibitem{Shaw2010}
D.~E. Shaw, {\it et~al.\/}, {\it {Atomic-Level Characterization of the
  Structural Dynamics of Proteins}\/}, {\it Science\/} {\bf 330}, 341 (2010).

\bibitem{SchlickBook}
T.~Schlick, {\it Molecular Modeling and Simulation: An Interdisciplinary
  Guide\/} (Springer-Verlag, 2002).

\bibitem{KarplusNobel}
M.~Karplus, {\it Development of Multiscale Models for Complex Chemical Systems:
  From $\text{H}+\text{H}_2$ to Biomolecules (Nobel Lecture)\/}, {\it Angew.
  Chem. Int. Ed.\/} {\bf 53}, 9992 (2014).

\bibitem{Amber_2}
D.~A. Case, {\it et~al.\/}, {\it The Amber biomolecular simulation programs\/},
  {\it J. Comput. Chem.\/} {\bf 26}, 1668 (2005).

\bibitem{CHARMM}
B.~R. Brooks, {\it et~al.\/}, {\it CHARMM: The biomolecular simulation
  program\/}, {\it J. Comput. Chem.\/} {\bf 30}, 1545 (2009).

\bibitem{GROMACS_1}
H.~Berendsen, D.~{van der Spoel}, R.~{van Drunen}, {\it GROMACS: A
  message-passing parallel molecular dynamics implementation\/}, {\it Comput.
  Phys. Commun.\/} {\bf 91}, 43 (1995).

\bibitem{LAMMPS}
A.~P. Thompson, {\it et~al.\/}, {\it {LAMMPS} - a flexible simulation tool for
  particle-based materials modeling at the atomic, meso, and continuum
  scales\/}, {\it Comput. Phys. Commun.\/} {\bf 271}, 108171 (2022).

\bibitem{NAMD}
J.~C. Phillips, {\it et~al.\/}, {\it Scalable molecular dynamics on CPU and GPU
  architectures with NAMD\/}, {\it J. Chem. Phys.\/} {\bf 153}, 044130 (2020).

\bibitem{Hammonds2020}
K.~D. Hammonds, D.~M. Heyes, {\it Shadow Hamiltonian in classical NVE molecular
  dynamics simulations: A path to long time stability\/}, {\it J. Chem.
  Phys.\/} {\bf 152}, 024114 (2020).

\bibitem{Wong2016}
J.~Wong-ekkabut, M.~Karttunen, {\it The good, the bad and the user in soft
  matter simulations\/}, {\it Biochim. Biophys. Acta - Biomembr.\/} {\bf 1858},
  2529 (2016).

\bibitem{Hockney1988}
R.~W. {Hockney}, J.~W. {Eastwood}, {\it Computer Simulation Using Particles\/}
  (CRC Press, 1988).

\bibitem{Darden1993}
T.~Darden, D.~York, L.~Pedersen, {\it Particle mesh Ewald: An $N\cdot\log(N)$
  method for Ewald sums in large systems\/}, {\it J. Chem. Phys.\/} {\bf 98},
  10089 (1993).

\bibitem{EssmannPedersen1995}
U.~Essmann, {\it et~al.\/}, {\it {A smooth particle mesh Ewald method}\/}, {\it
  J. Chem. Phys.\/} {\bf 103}, 8577 (1995).

\bibitem{Petersen1995}
H.~G. Petersen, {\it {Accuracy and efficiency of the particle mesh Ewald
  method}\/}, {\it J. Chem. Phys.\/} {\bf 103}, 3668 (1995).

\bibitem{Kohnke2020}
B.~Kohnke, C.~Kutzner, H.~Grubmüller, {\it A GPU-Accelerated Fast Multipole
  Method for GROMACS: Performance and Accuracy\/}, {\it J. Chem. Theory
  Comput.\/} {\bf 16}, 6938 (2020).

\bibitem{Wennberg2013}
C.~L. Wennberg, T.~Murtola, B.~Hess, E.~Lindahl, {\it Lennard-Jones Lattice
  Summation in Bilayer Simulations Has Critical Effects on Surface Tension and
  Lipid Properties\/}, {\it J. Chem. Theory Comput.\/} {\bf 9}, 3527 (2013).

\bibitem{Davis1984}
M.~H.~A. Davis, {\it {Piecewise-Deterministic Markov Processes: A General Class
  of Non-Diffusion Stochastic Models}\/}, {\it J. R. Stat. Soc. Series B Stat.
  Methodol.\/} {\bf 46}, 353 (1984).

\bibitem{Bernard2011}
E.~P. Bernard, W.~Krauth, {\it {Two-Step Melting in Two Dimensions: First-Order
  Liquid-Hexatic Transition}\/}, {\it Phys. Rev. Lett.\/} {\bf 107}, 155704
  (2011).

\bibitem{Michel2014JCP}
M.~{Michel}, S.~C. {Kapfer}, W.~{Krauth}, {\it {Generalized event-chain Monte
  Carlo: Constructing rejection-free global-balance algorithms from
  infinitesimal steps}\/}, {\it J. Chem. Phys.\/} {\bf 140}, 054116 (2014).

\bibitem{WuTepperVoth2006}
Y.~Wu, H.~L. Tepper, G.~A. Voth, {\it {Flexible simple point-charge water model
  with improved liquid-state properties}\/}, {\it J. Chem. Phys.\/} {\bf 124},
  024503 (2006).

\bibitem{deLeeuw1980-1}
S.~W. de~Leeuw, J.~W. Perram, E.~R. Smith, {\it {Simulation of electrostatic
  systems in periodic boundary conditions. I. Lattice sums and dielectric
  constants}\/}, {\it Proc. R. Soc. A\/} {\bf 373}, 27 (1980).

\bibitem{Bernard2009}
E.~P. Bernard, W.~Krauth, D.~B. Wilson, {\it {Event-chain Monte Carlo
  algorithms for hard-sphere systems}\/}, {\it Phys. Rev. E\/} {\bf 80}, 056704
  (2009).

\bibitem{Krauth2021eventchain}
W.~Krauth, {\it {Event-Chain Monte Carlo: Foundations, Applications, and
  Prospects}\/}, {\it Front. Phys.\/} {\bf 9}, 229 (2021).

\bibitem{Metropolis1953}
N.~{Metropolis}, A.~W. {Rosenbluth}, M.~N. {Rosenbluth}, A.~H. {Teller},
  E.~{Teller}, {\it {Equation of State Calculations by Fast Computing
  Machines}\/}, {\it J. Chem. Phys.\/} {\bf 21}, 1087 (1953).

\bibitem{GammaDesignPatterns1994}
E.~Gamma, R.~Helm, R.~Johnson, J.~M. Vlissides, {\it {Design Patterns: Elements
  of Reusable Object-Oriented Software}\/} (Addison-Wesley Professional, 1994).

\bibitem{Harland2017}
J.~Harland, M.~Michel, T.~A. Kampmann, J.~Kierfeld, {\it {Event-chain Monte
  Carlo algorithms for three- and many-particle interactions}\/}, {\it EPL\/}
  {\bf 117}, 30001 (2017).

\bibitem{Faulkner2018}
M.~F. Faulkner, L.~Qin, A.~C. Maggs, W.~Krauth, {\it {All-atom computations
  with irreversible Markov chains}\/}, {\it J. Chem. Phys.\/} {\bf 149}, 064113
  (2018).

\bibitem{Klement2019}
M.~Klement, M.~Engel, {\it {Efficient equilibration of hard spheres with
  Newtonian event chains}\/}, {\it J. Chem. Phys.\/} {\bf 150}, 174108 (2019).

\bibitem{Michel2020}
M.~Michel, A.~Durmus, S.~S{\'{e}}n{\'{e}}cal, {\it {Forward Event-Chain Monte
  Carlo: Fast Sampling by Randomness Control in Irreversible Markov Chains}\/},
  {\it J. Comput. Graph. Stat.\/} {\bf 29}, 689 (2020).

\bibitem{Hoellmer2022Dipoles}
P.~Höllmer, A.~C. Maggs, W.~Krauth, {\it Hard-disk dipoles and non-reversible
  Markov chains\/}, {\it J. Chem. Phys.\/} {\bf 156}, 084108 (2022).

\bibitem{Hoellmer2020}
P.~H\"{o}llmer, L.~Qin, M.~F. Faulkner, A.~Maggs, W.~Krauth, {\it
  {JeLLyFysh-Version1.0~{\textemdash} a Python application for all-atom
  event-chain Monte Carlo}\/}, {\it Comput. Phys. Commun.\/} {\bf 253}, 107168
  (2020).

\bibitem{Li2020}
B.~Li, S.~Todo, A.~Maggs, W.~Krauth, {\it {Multithreaded event-chain Monte
  Carlo with local times}\/}, {\it Comput. Phys. Commun.\/} {\bf 261}, 107702
  (2021).

\bibitem{KapferKrauth2016}
S.~C. Kapfer, W.~Krauth, {\it {Cell-veto Monte Carlo algorithm for long-range
  systems}\/}, {\it Phys. Rev. E\/} {\bf 94}, 031302 (2016).

\bibitem{Walker1977AnEfficientMethod}
A.~J. Walker, {\it {An Efficient Method for Generating Discrete Random
  Variables with General Distributions}\/}, {\it ACM Trans. Math. Softw.\/}
  {\bf 3}, 253 (1977).

\bibitem{LewisShedler1979}
P.~A.~W. Lewis, G.~S. Shedler, {\it {Simulation of nonhomogeneous Poisson
  processes by thinning}\/}, {\it Nav. Res. Logist. Q.\/} {\bf 26}, 403 (1979).

\bibitem{Saunders2021}
W.~R. Saunders, J.~Grant, E.~H. Müller, {\it A new algorithm for electrostatic
  interactions in Monte Carlo simulations of charged particles\/}, {\it J.
  Comput. Phys.\/} {\bf 430}, 110099 (2021).

\bibitem{greenhard1987}
L.~Greengard, V.~Rokhlin, {\it {A fast algorithm for particle simulations}\/},
  {\it J. Comput. Phys.\/} {\bf 73}, 325 (1987).

\bibitem{DLMONTE}
J.~Purton, J.~Crabtree, S.~Parker, {\it DL\_MONTE: a general purpose program
  for parallel Monte Carlo simulation\/}, {\it Mol. Simul.\/} {\bf 39}, 1240
  (2013).

\bibitem{Kolafa1992}
J.~Kolafa, J.~W. Perram, {\it Cutoff Errors in the Ewald Summation Formulae for
  Point Charge Systems\/}, {\it Mol. Simul.\/} {\bf 9}, 351 (1992).

\bibitem{Deserno1998}
M.~Deserno, C.~Holm, {\it {How to mesh up Ewald sums. II. An accurate error
  estimate for the particle–particle–particle-mesh algorithm}\/}, {\it J.
  Chem. Phys.\/} {\bf 109}, 7694 (1998).

\bibitem{Vanetti2018}
P.~Vanetti, A.~Bouchard-Côté, G.~Deligiannidis, A.~Doucet, {\it
  Piecewise-Deterministic Markov Chain Monte Carlo\/}, {\it arXiv: 1707.05296
  [stat.ME]\/}  (2018).

\bibitem{Fearnhead2018}
P.~Fearnhead, J.~Bierkens, M.~Pollock, G.~O. Roberts, {\it {Piecewise
  Deterministic Markov Processes for Continuous-Time Monte Carlo}\/}, {\it
  Statist. Sci.\/} {\bf 33}, 386  (2018).

\bibitem{Bierkens2020}
J.~Bierkens, S.~Grazzi, K.~Kamatani, G.~Roberts, {\it Proceedings of the 37th
  International Conference on Machine Learning\/}, H.~Daum\'e~III, A.~Singh,
  eds. (PMLR, 2020), vol. 119 of {\it Proceedings of Machine Learning
  Research\/}, pp. 908--918.

\bibitem{LAMMPS_V}
S.~Plimpton, A.~Thompson, S.~Moore, A.~Kohlmeyer, R.~Berger, {\it LAMMPS,
Feature release 8 February 2023\/}, \url{https://www.lammps.org} (2023). Last
accessed: 31 May 2023.

\bibitem{Shinoda2004}
W.~Shinoda, M.~Shiga, M.~Mikami, {\it Rapid estimation of elastic constants by
molecular dynamics simulation under constant stress\/}, {\it Phys. Rev. B\/}
{\bf 69}, 134103 (2004).

\bibitem{Tuckerman2006}
M.~E. Tuckerman, J.~Alejandre, R.~López-Rendón, A.~L. Jochim, G.~J. Martyna,
{\it A Liouville-operator derived measure-preserving integrator for molecular
dynamics simulations in the isothermal–isobaric ensemble\/}, {\it J. Phys.
A: Math. Gen.\/} {\bf 39}, 5629 (2006).

\bibitem{DLMONTE_V}
A.~Brukhno, {\it et~al.\/}, {\it DL\_MONTE, version 2.07\/},
\url{https://gitlab.com/dl_monte/user-hub/-/wikis/home} (2020). Last
accessed: 31 May 2023.

\bibitem{DLMONTE2}
A.~V. Brukhno, {\it et~al.\/}, {\it DL\_MONTE: a multipurpose code for Monte
Carlo simulation\/}, {\it Mol. Simul.\/} {\bf 47}, 131 (2021).

\bibitem{Ewald1921}
P.~P. Ewald, {\it Die Berechnung optischer und elektrostatischer
Gitterpotentiale\/}, {\it Ann. Phys.\/} {\bf 369}, 253 (1921).

\bibitem{JF_V}
P.~H\"ollmer, A.~C. Maggs, W.~Krauth, {\it JeLLyFysh, version 2.0\/},
\url{https://github.com/jellyfysh/JeLLyFysh} (2023). Last accessed: 01 June
2023.

\bibitem{Qin2020Thesis}
L.~Qin, {\it {Application of irreversible Monte Carlo in realistic long-range
systems}\/}, Phd thesis, {Universit{\'e} Paris Sciences et Lettres} (2020).

\bibitem{HoellmerBoroczky2022}
P.~H\"{o}llmer, N.~Noirault, B.~Li, A.~C. Maggs, W.~Krauth, {\it {Sparse
Hard-Disk Packings and Local Markov Chains}\/}, {\it J. Stat. Phys.\/} {\bf
187} (2022).

\bibitem{Playmol_V}
C.~R.~A. Abreu, {\it Playmol, commit \texttt{67eb56c} from 26 November 2019\/},
\url{http://atoms.peq.coppe.ufrj.br/playmol/index.html} (2019). Last
accessed: 01 June 2023.

\bibitem{Michel2023}
A.~Monemvassitis, A.~Guillin, M.~Michel, {\it PDMP Characterisation of
Event-Chain Monte Carlo Algorithms for Particle Systems\/}, {\it J. Stat.
Phys.\/} {\bf 190}, 66 (2023).

\bibitem{Chen1999}
F.~Chen, L.~Lovász, I.~Pak, {\it {Lifting Markov Chains to Speed up
Mixing}\/}, {\it Proceedings of the 17th Annual ACM Symposium on Theory of
Computing\/} p. 275 (1999).

\bibitem{Diaconis2000}
P.~Diaconis, S.~Holmes, R.~M. Neal, {\it {Analysis of a nonreversible Markov
chain sampler}\/}, {\it Ann. Appl. Probab.\/} {\bf 10}, 726 (2000).

\bibitem{Qin2022}
L.~Qin, P.~Höllmer, W.~Krauth, {\it Direction-sweep Markov chains\/}, {\it J.
Phys. A: Math. Theor.\/} {\bf 55}, 105003 (2022).

\bibitem{Peters_2012}
E.~A. J.~F. Peters, G.~de~With, {\it {Rejection-free Monte Carlo sampling for
general potentials}\/}, {\it Phys. Rev. E\/} {\bf 85}, 026703 (2012).

\bibitem{RobertsFibonacci}
M.~Roberts, {\it How to evenly distribute points on a sphere more effectively
than the canonical Fibonacci Lattice\/},
\url{https://web.archive.org/web/20230527141637/http://extremelearning.com.au/how-to-evenly-distribute-points-on-a-sphere-more-effectively-than-the-canonical-fibonacci-lattice/}
(2020). Last accessed: 30 May 2023.

\bibitem{FengFibonacci}
Y.~Feng, {\it An effective energy-conserving contact modelling strategy for
spherical harmonic particles represented by surface triangular meshes with
automatic simplification\/}, {\it Comput. Methods Appl. Mech. Eng.\/} {\bf
379}, 113750 (2021).

\bibitem{Saff2016}
D.~P. Hardin, T.~J. Michaels, E.~B. Saff, {\it A Comparison of Popular Point
Configurations on $\mathbb{S}^2$\/}, {\it arXiv: 1607.04590 [math.NA]\/}
(2016).

\bibitem{Kleinert2015}
B.~Keinert, M.~Innmann, M.~S\"{a}nger, M.~Stamminger, {\it Spherical Fibonacci
Mapping\/}, {\it ACM Trans. Graph.\/} {\bf 34}, 193 (2015).

\end{thebibliography}

\bibliographystyle{ScienceWithTitle}

\paragraph*{Acknowledgments}\mbox{}\\
We thank L. Qin for helpful discussions.\\
\textbf{Funding:}
P.H. acknowledges support from the Studienstiftung des deutschen Volkes and
from Institut Philippe Meyer. W.K. acknowledges support from the Alexander von
Humboldt Foundation.\\
\textbf{Author contributions:} \\
Conceptualization: PH, ACM, WK \\
Formal Analysis: PH \\
Investigation: PH, ACM \\
Methodology: PH, ACM, WK \\
Project administration: ACM, WK \\
Software: PH \\
Supervision: ACM, WK \\
Validation: PH, ACM, WK \\
Visualization: PH \\
Writing---original draft: PH, WK \\
Writing---review \& editing: PH, ACM, WK \\
\textbf{Competing interests:} The authors declare that they have no competing 
interests. \\
\textbf{Data and materials availability:} The \JF software is made
available under the GNU GPLv3 license at \url{https://github.com/jellyfysh}. 
All configuration files for the simulations of this paper are part of \JF.

\paragraph*{Supplementary Materials} \mbox{} \\
Materials and Methods\\
Supplementary Text\\
Fig.~S1\\
References (45--63)

\newpage

\end{document}


	
	\baselineskip24pt

	\maketitle 
	

\noindent
\textbf{The PDF file includes:}\mbox{}\\
Materials and Methods\\
Supplementary Text\\
Fig.~S1

\paragraph*{Materials and Methods} \label{sec:materials} \mbox{} \\
In this section, we describe the SPC/Fw water model that we studied in this 
paper. We further specify the simulation protocols for the different sampling 
algorithms. Finally, we report on our analysis procedure of the time series of 
the electric polarization.

\subparagraph*{SPC/Fw Water Model}\mbox{}\\
The atomistic flexible simple point-charge water model SPC/Fw defines a
water molecule by three charged interaction sites that represent the oxygen and
hydrogen atoms~(\emph{21}). We treat the canonical ensemble in a
cubic box with periodic boundary conditions, i.e., $N$ water molecules with
$N_a=3N$ atoms in a periodically repeated cubic box of side length $L$ at a
given temperature $T\sim\SI{300}{\kelvin}$. Within any water molecule 
$i\in\{1,\ldots,N\}$, the
intramolecular interactions of the SPC/Fw model consist of two harmonic bond
potentials $U_{\text{bond}}^{i,1}$ and $U_{\text{bond}}^{i,2}$ that lead to
fluctuations of the \ch{O-H} bond lengths around their equilibrium length.
Likewise, a harmonic bending potential $U^{i}_\text{bend}$ yields a fluctuation
of the \ch{H-O-H} opening angle around an equilibrium value. The intermolecular
interactions between two different water molecules $i$ and $j$ consist of a
Lennard-Jones potential $U^{ij}_\text{LJ}$ between the two oxygen atoms, and a
Coulomb potential $U^{ij}_\text{C}$ between all nine pairs of charged atoms.
The intermolecular potentials explicitly include the interactions between all 
periodic
images of the two involved water molecules. The total potential $U$ of an
all-atom configuration $\xvec$ is a sum of these factor potentials:
\begin{equation}
	\label{eq:fac}
	U(\xvec) = \sum_{i=1}^{N} \left[ U_{\text{bond}}^{i,1}(\xvec) + 
	U_{\text{bond}}^{i,2}(\xvec) +
	U_{\text{bend}}^i(\xvec) \right] + \sum_{i=1}^N \sum_{j=1}^{i-1} 
	\left[ U_\text{LJ}^{ij}(\xvec) 
	+ U_\text{C}^{ij}(\xvec) \right].
\end{equation}
The bond and Lennard-Jones factor potentials depend on two atomic positions; 
the bending and Coulomb factor potentials depend on three and six atomic 
positions, respectively [see also Ref.~(\emph{28, Section V A})]. The 
empirical parameters of the different potentials
of the SPC/Fw water model are as in Ref.~(\emph{21}).

\subparagraph*{Simulation Protocols---Molecular Dynamics}\mbox{}\\
We use the feature release from February 8, 2023, of \LA~(\emph{45, 8})
for the molecular-dynamics simulations of the SPC/Fw water model in this paper.
We use a spherical cut-off for the Lennard-Jones potential. The Coulomb 
potential is treated by a particle--particle
particle--mesh solver~(\emph{12}). Unless explicitly specified otherwise
(as, e.g., in Fig.~4B), the solver uses a target accuracy of $10^{-6}$. 
Coulomb interactions are partly treated in
discretized reciprocal space, whereby the grid size is chosen to meet the target
accuracy based on analytic error estimates obtained from a specific global 
charge distribution~(\emph{40, 15, 41}). Thermostatting 
is
achieved by integrating Nos\'e--Hoover-chain equations of motion for the
canonical ensemble~(\emph{46}) with a time-reversible measure-preserving
Verlet integrator~(\emph{47}). The time step is
$\SI{1}{\femto\second}$, and the temperature is relaxed in a timespan of 
roughly $300$ time steps. This simulation protocol for molecular dynamics 
in \LA is validated by comparing numerical results for $N=2$ SPC/Fw water 
molecules of all sampling methods of this paper (see \fig{fig:Validation}).


\subparagraph*{Simulation Protocols---Metropolis Algorithm}\mbox{}\\
We use version $2.07$ of the \textsc{DL\_MONTE} software
package~(\emph{48, 39, 49}) to sample the 
canonical ensemble of
the SPC/Fw water model with the reversible Metropolis
algorithm~(\emph{25}), and minimally amended the software to output
the electric polarization. We use a spherical cut-off for the Lennard-Jones
potential. The Coulomb interaction is treated by an Ewald
summation~(\emph{50}) with a target-accuracy tolerance of $10^{-6}$. The
simulations with the event-chain Monte Carlo algorithm in this paper move a
single atom at a time. In order to directly compare the dynamics of the
reversible Metropolis algorithm with the non-reversible event-chain Monte Carlo
algorithm (see Fig.~3C), each trial proposes a new position of a single random
atom in a proposal cube around its original position. The new position is
accepted with the usual Metropolis criterion based on the change of energy. The
size of the proposal cube is separately adapted for the hydrogen and oxygen
atoms during the simulation to obtain a target acceptance rate of
$\SI{37}{\percent}$. This simulation protocol for the Metropolis algorithm in 
\textsc{DL\_MONTE } is validated by comparing numerical results for $N=2$ 
SPC/Fw water 
molecules of all sampling methods of this paper (see \fig{fig:Validation}).

\subparagraph*{Simulation Protocols---Event-Chain Monte Carlo}\mbox{}\\
We develop version $2.0$ of \JF~(\emph{51, 32}) for the
event-chain Monte Carlo simulations of the SPC/Fw water model in this paper. In
addition to the straight variant implemented in versions $< 2.0$, version
2.0 implements the generalized Newtonian event-chain Monte Carlo variant (see
\nameref{sec:sup} for details). We use the factorization in \eq{eq:fac}. In
particular, we group all the charge--charge Coulomb interactions
between two water molecules into a single factor. Theory then predicts an
optimal logarithmic
increase of the number of events from these molecular Coulomb
factors~(\emph{28}) (that is numerically confirmed in Fig.~3B). 

Straight event-chain Monte Carlo (used in Fig.~3C)
periodically aligns the unit velocity $|\vvec_a|= 1$ of the active atom with the
coordinate axes. We change the velocity after a chain time of
$\tau_\text{chain}=0.2N$~(\emph{52}). In an event triggered by a
factor $M$, the
velocity is transferred to another atom which belongs to $M$. For factors
with more than two atoms, the next active atom is chosen according to the ratio
lifting
scheme~(\emph{27, 28}). 

Newtonian event-chain Monte Carlo assigns a velocity label to every atom, but
only moves a single one at any time. The velocities are initialized so that the
average speed of the active atom during the simulation is approximately one:
$\langle |\vvec_a| \rangle \approx 1$ (see \nameref{sec:sup} for details). This
is possible because the velocities have no kinematic meaning as they have in
molecular dynamics. The physical temperature only rescales the event rates [see
\eq{eq:event} in the \nameref{sec:sup}]. The velocities are resampled in large
time intervals of $\tau_\text{chain}=\num{10000}{N}$. This is because Newtonian
event-chain Monte Carlo rotates dipoles most efficiently in the limit
$\tau_\text{chain}\rightarrow\infty$, while a finite value of
$\tau_\text{chain}$ ensures
irreducibility~(\emph{53, 31}). Although our
generalized Newtonian event-chain Monte Carlo can consider a general mass matrix
of the atoms in its events, we set the masses of all atoms to be equal. Then,
the events of factors $M$ of two atoms can be treated in a deterministic way in
which the two involved velocities and the active atom always change. For factors
$M$ of more than two atoms, we apply a stochastic scheme that corresponds to a
generalized ratio lifting scheme (see the Newtonian-pair and Newtonian-general 
schemes in the \nameref{sec:sup}). 

The events of the bond factors are computed exactly [see
\eq{eq:Poisson} in the \nameref{sec:sup}], while the bending
factors are treated with a piecewise-linear bounding potential followed by a
Poisson thinning procedure [see~(\emph{32, Section 4.4.5})].
The cell-veto algorithm~(\emph{34}) bundles most molecular Coulomb
factors and relies on a
cell-occupancy system that tracks the molecular barycenters [we generally adapt
the setup described in Ref.~(\emph{32, Section 5.3.4})]. Very close
molecule pairs or surplus molecules in the cell-occupancy system are separately
treated with a piecewise-linear bounding potential. The cell bounds for
$10$ different Fibonacci vectors (see \nameref{sec:sup} for details) are
estimated by varying the position of an atom in one cell, and the position of a
dipole in the other cell. Here, the dipole is aligned with the direction of the
gradient of the charge--dipole Coulomb interaction. The maximum event rate
multiplied by an empirical prefactor yields the cell bound. During the
simulation, the cell bounds in the Walker table are
rescaled to the actual
speed and charge of the active atom. Events from the cell-veto algorithm are
confirmed with a Poisson thinning procedure that compares the actual event rate
of the Coulomb interaction between two molecules with the sampled cell bound. To
compute the real event rate, we compute the gradients of the Coulomb potential
between the nine pairs of charged atoms [see also \eq{eq:event} in the
\nameref{sec:sup}] with an Ewald summation~(\emph{50}) that is tuned to
machine precision without any assumption on the global charge distribution. We
also use the cell-veto algorithm to treat most Lennard-Jones factors, although
we cut off the interaction beyond the closest images for consistency with the
other sampling algorithms. We carefully checked that an alternative spherical 
cut-off does 
not change the results of this paper.
The cell-veto algorithm for the Lennard-Jones interaction relies on a 
cell-occupancy system that tracks
the oxygens. Very close or surplus oxygen pairs are treated directly [see
\eq{eq:Poisson} in the \nameref{sec:sup}]. The
cell bounds for $10$ different Fibonacci vectors are estimated by varying the
oxygen positions evenly in the two cells, and again including an empirical
prefactor.
The parameters of the cell-occupancy systems, as the cell sizes and the number
of nearby cells that are excluded from the cell-veto algorithm, are tuned to
minimize computer time. As a rule of thumb, the largest possible cells that do 
not yield any surplus water molecules, and two excluded layers are found to be 
a good choice. 

These simulation protocols for straight and Newtonian 
event-chain Monte Carlo
in \JF are validated by comparing numerical results for $N=2$ SPC/Fw water 
molecules of all sampling methods of this paper (see \fig{fig:Validation}).

\subparagraph*{Simulation Protocols---Creation of Initial 
Configurations}\mbox{}\\
We create initial configurations in the range from  $N=64$ to $N=2744$ SPC/Fw
water molecules with hydrogen mass $m_H = \SI{1.0079}{\dalton}$ and oxygen mass 
$m_O = \SI{15.9994001}{\dalton}$ at a density of
$\rho=\SI{0.97}{\gram\per\centi\meter\cubed}$
using the software package \textsc{Playmol}~(\emph{54}) (commit
\texttt{67eb56c}
from 26 November 2019). Initial configurations are equilibrated using a
molecular-dynamics
simulation of \LA in the isothermal--isobaric ensemble at pressure 
$P=\SI{1}{\atm}$. Similar to thermostatting, 
barostatting is achieved by integrating the appropriate Nos\'e--Hoover-chain 
equations of motion with a Verlet integrator~(\emph{46, 47}) 
with a time step of $\SI{1}{\femto\second}$.  
This results in initial configurations for the simulations in the canonical
ensemble with the different sampling algorithms at slightly different densities 
for different $N$. We confirmed that these small density variations do not 
influence the results of this paper.

\subparagraph*{Analysis of the Electric Polarization}\mbox{}\\
We sample the electric polarization $\mathbf{P}$ (or the total electric dipole
moment $\mathbf{P}=\sum_{i=1}^N \mathbf{p}_i$, where $\mathbf{p}_i$ is the
molecular dipole moment of the water molecule $i$) of the SPC/Fw water system
with the different sampling algorithms. The generated \quot{time}-series
$\mathbf{P}(t)$ (where the \quot{time} $t$ only has a a physical meaning in
molecular dynamics) yields the normalized autocorrelation function
\begin{equation} \Phi_\mathbf{P}(\tau) = \frac{\langle
	\mathbf{P}(t)\cdot\mathbf{P}(t + \tau)\rangle_t}{\langle
	\mathbf{P}(t)\cdot\mathbf{P}(t)\rangle_t}, 
\end{equation}
where we explicitly
assume that $\langle \mathbf{P} \rangle = 0$ for $t\rightarrow\infty$. The 
first part of each trajectory in the canonical ensemble is not considered in
the equilibrium average. We
observe an exponential decay in all sampling algorithms and fit an exponential
$\sim\exp(-\tau/\tau_\Phi)$ to extract the time constant $\tau_\Phi$ and its 
standard error [see also Ref.~(\emph{21})]. We na\"ively 
parallelize the Metropolis and event-chain 
Monte 
Carlo algorithms by running $\SIrange{20}{50}{}$ independent simulations (where 
the 
number of runs grows with $N$). For every $\tau$, we then compute the median 
autocorrelation function $\Phi_\mathbf{P}(\tau)$ and estimate its error with 
the bootstrap method. These errors are then considered in the least-squares fit.

The units of the autocorrelation time $\tau_\Phi$ for the different sampling
algorithms are physical time for molecular dynamics, Monte-Carlo trials for the
Metropolis algorithm, and continuous Monte-Carlo time for the event-chain Monte
Carlo algorithms. As a first measure of efficiency of the different dynamics, we
compare the autocorrelation distances $d_\Phi$ (see Fig.~3C). It gives the
average cumulative distance moved by the atoms in each of the sampling schemes.
Since only a single atom moves in our implementation of the event-chain Monte
Carlo algorithm, it follows that
$d_\Phi^\text{ECMC}=\langle|\mathbf{v}_a|\rangle^\text{ECMC} \,
\tau_\Phi^\text{ECMC}$, where
$\langle|\mathbf{v}_a|\rangle^\text{ECMC}$ is the average speed of the active 
atom during
the simulation. For molecular dynamics, we can compute the average speed of 
the hydrogen and oxygen atoms $t\in\{\ch{H}, \ch{O}\}$ with the 
Maxwell--Boltzmann distribution:
\begin{equation}
	\langle \left| \vvec_t \right| \rangle^\text{MD} = \sqrt{\frac{8 
	k_\mathrm{B} T}{\pi 
	m_t}},
\end{equation}
which yields $d_\Phi^\text{MD}=N (2 \langle |\vvec_{\ch{H}}| \rangle^\text{MD} 
+ \langle | \vvec_{\ch{O}} | \rangle^\text{MD}) \, \tau_\Phi^\text{MD}$. In the 
Metropolis algorithm, each trial samples a random displacement vector 
$\mathbf{d}_t=[\ran(-\delta_t, \delta_t), \ran(-\delta_t, \delta_t), 
\ran(-\delta_t, \delta_t)]^T$ for a random atom, where $\delta_t$ is tuned 
independently for the 
oxygen and hydrogen atoms $t\in\{ \ch{H}, \ch{O} \}$ to accept the displacement 
with probability $p\approx0.37$. We can compute the average length $\langle 
|\mathbf{d}_t|\rangle^\text{Met}$ of the sampled displacement vector as 
\begin{equation}
	\langle |\mathbf{d}_t|\rangle^\text{Met} = 
	\frac{1}{8\delta_t^3}\int_{-\delta_t}^{\delta_t}\mathrm{d}x\,
	\int_{-\delta_t}^{\delta_t}\mathrm{d}y\,
	\int_{-\delta_t}^{\delta_t}\mathrm{d}z\, \sqrt{x^2 + y^2 + z^2} \approx 
	\num{0.961}\,\delta_t.
\end{equation}
Neglecting correlations between rejection probability and the displacement for
this first measure of efficiency, we get
$d_\Phi^\text{Met}=p\,(2\langle|\mathbf{d}_{\ch{H}}|\rangle^\text{Met}/3 + 
\langle|\mathbf{d}_{\ch{O}}|\rangle^\text{Met}/3)\,\tau_\Phi^\text{Met}$.

Finally, the computer autocorrelation time of our generalized Newtonian
event-chain Monte Carlo algorithm (see Fig.~4B) is obtained by combining the
autocorrelation distance (see Fig.~3C), the event rates (see Fig.~3B), and the
computer time per event in \JF (see Fig.~3A). Similarly, we measure the computer
time per time step of molecular dynamics in \LA for different target accuracies
of its particle--particle particle--mesh solver (see Fig.~4A). This is then 
combined with the autocorrelation time
in the number of time steps.

\paragraph*{Supplementary Text} \label{sec:sup} \mbox{} \\
In this supplementary text, we generalize Newtonian
event-chain Monte Carlo to smooth factor potentials depending on an arbitrary 
number of atoms. We further prove that this algorithm satisfies the necessary 
global-balance condition. Finally, we report on our generation of Fibonacci 
vectors on the unit sphere. 

\subparagraph*{Generalized Newtonian Event-Chain Monte Carlo} \mbox{} \\
Newtonian event-chain Monte Carlo was originally formulated for the hard-sphere
model that contains peculiar stepwise-changing two-body factor
potentials~(\emph{29}). Among various event-chain Monte Carlo variants,
it was shown to escape faster from sparse hard-disk
packings~(\emph{53}) and to produce favorable rotation dynamics 
in tethered
hard-disk dipoles~(\emph{31}). Its superiority is confirmed for
the
rotation dynamics of SPC/Fw water molecules in this paper, but requires its
generalization to smooth factor potentials that depend on an arbitrary number of
atom positions.

We follow the initial introduction of non-reversible event-chain Monte Carlo as
a lifted continuous-time Markov process~(\emph{20})
[see also the formulation as a piecewise-deterministic Markov process
in Ref.~(\emph{55})]. For Newtonian
event-chain Monte Carlo for $N_a$ three-dimensional atoms, the lifting
framework~(\emph{56, 57}) [see also
Refs~(\emph{24}) and (\emph{58, Appendix A})] extends the
physical all-atom configuration $\xvec$ to $(\xvec, \vvec, i)$, with auxiliary 
lifting
variables that represent an all-atom velocity $\vvec$ and an activity label $i$,
respectively. The physical sample space $\xvec\in\Omega$ is thus extended to 
the lifted
sample space $(\xvec, \vvec, i) \in \widehat{\Omega} = \Omega \times
\mathcal{V}^{3N_a}(E_\text{kin}) \times \mathcal{N}$, where
$\mathcal{N}=\{1\ldots N_a\}$ the set of atom indices, and $\mathcal{V}^{3N_a}
(E_\text{kin}) = \{ \vvec \in \mathbb{R}^{3N_a}: \vvec^T \cdot
\mathbf{M} \cdot \vvec = 2E_\text{kin} \}$ with a positive-definite
symmetric mass matrix $\mathbf{M}$ and total kinetic energy
$E_\text{kin}$. The Markov process targets the lifted stationary distribution
$\hat{\pi}(\xvec,\vvec,i) = \pi(\xvec) \times \mu_\mathcal{V}(\vvec) \times
\mu_\mathcal{N}(i)$ that separates into the factorized Boltzmann distribution 
$\pi(\xvec)
\propto \exp[-\beta U(\xvec)]=\prod_M \exp[-\beta U_M(\xvec)]$, and the uniform 
distributions
$\mu_\mathcal{V}(\vvec)$ on $\mathcal{V}^{3N_a}$, and $\mu_\mathcal{N}(i)$
on $\mathcal{N}$.

Given a lifted configuration $(\xvec(t_0), \vvec(t_0), i)$ at time $t_0$,
Newtonian event-chain Monte Carlo continuously moves the single active atom $i$
starting from its position $\xvec_i(t_0)$ with its constant velocity $\vvec_i =
\vvec_i(t_0)$ up to  an event at time $t_\text{ev}>t_0$, which interrupts the
motion:
$\xvec_i(t) =
\xvec_i(t_0) + \vvec_i \, (t - t_0)$ for $t_0 <= t < t_\text{ev}$. The
velocities $\vvec_j$ of all other atoms $j\neq i$ in the all-atom velocity
$\vvec$ are, at this point, hypothetical, that is, mere labels. The
time-dependent event rate
$\lambda(t)$, that is, the probability density to interrupt the
piecewise-deterministic motion of the active atom $i$, is given by a sum of 
factor event rates $\lambda_M(t)$:
\begin{equation} 
\label{eq:event}
\lambda(t) = \sum_M \lambda_M(t) = \sum_{M} \beta
\max\left[0,\grad_{\xvec_i} U_M\left(\xvec(t)\right)\cdot\vvec_i\right].
\end{equation} 
Every factor $M$ can be considered as statistically independent. Each of them
stochastically
generates a candidate event time $t_{\text{ev},M}$ in an inhomogeneous Poisson
process based on its factor event rate $\lambda_M(t)\geq 0$ (that is only
nonzero for factor potentials $U_M$ that are actually changed by the motion of
the active atom):
\begin{equation}
\label{eq:Poisson}
	\ran_M(0,1) = \exp\left[- \int_{t_0}^{t_{\text{ev}, M}}\lambda_M(t)\, 
	\mathrm{d}t\right].
\end{equation}
Here, $\ran_M(0,1)$ is a uniformly distributed random number between $0$ and $1$
that is drawn separately for each factor $M$. In \eq{eq:Poisson}, the cumulative
increments of the factor potential $U_M$ under the motion of the active atom
with its velocity $\vel_i$ since $t_0$ are equal to an exponentially 
distributed random
number with mean $1/\beta$ at the time $t_{\text{ev},M}$.
If the computation of $t_{\text{ev}, M}$ from
\eq{eq:Poisson} is tedious or even impossible, a Poisson thinning procedure may 
yield candidate event times~(\emph{36}).  

At the event time $t_\text{ev}=\min_M t_{\text{ev}, M}$, the motion of the
active atom $i$ is interrupted at the lifted configuration $(\xvec =
\xvec(t_\text{ev}), \vvec,i)$. An event changes the lifting variables and sets
the initial lifted configuration $(\xvec, \vvec', j)$ for the next piece in the
piecewise-deterministic Markov process. The event is realized by a unique 
event-factor $M_\text{ev} = \argmin_\factor t_{\text{ev}, M}$. Generalized 
Newtonian 
event-chain
Monte Carlo proposes an update of the all-atom velocity $\vvec \rightarrow
\vvec'$ at time $t_\text{ev}$ with a force kick in the direction of the gradient
of the event-factor potential:
\begin{equation}
\label{eq:kick}
	\vvec' = \vvec - 2 \, \frac{\vvec^T \cdot \grad_{\xvec} 
	U_{M_\text{ev}}}{(\grad_{\xvec}U_{M_\text{ev}})^T \cdot 
	\mathbf{M}^{-1} \cdot \grad_{\xvec}U_{M_\text{ev}}} \, 
	\mathbf{M}^{-1}\cdot\grad_{\xvec} 
	U_{M_\text{ev}}.
\end{equation}
This force kick leaves the kinetic energy $2E_\text{kin} = \vvec^T \cdot
\mathbf{M} \cdot \vvec$ invariant and applying it twice yields the original 
$\vvec$~(\emph{59}). 
The event-factor potential only acts on a small subensemble $k\in M_\text{ev} $
of atoms, and thus only the velocities
$\vel_{k}$ of these contributing atoms are possibly modified. For the simple 
choice of an identity mass matrix $\mathbf{M} = \mathbf{I}$, \eq{eq:kick} 
may be written as
\begin{equation}
	\label{eq:kickdiagonal}
	\vel_k' = \vel_k - 2 \, \frac{\sum_{j\in M_\text{ev}} \vvec_j^T 
	\cdot \grad_{\xvec_j} U_{M_\text{ev}}}{\sum_{j\in M_\text{ev}} \left| 
	\grad_{\xvec_j} U_{M_\text{ev}} \right|^2}\, \grad_{\xvec_k} 
	U_{M_\text{ev}}.
\end{equation}

In Newtonian event-chain Monte Carlo, only a single active atom moves with 
its
velocity at any time. At an event, it must choose the next active atom
$k\in M_\text{ev}$ from the event factor. It also chooses to either change the 
velocity $\vvec \rightarrow \vvec'$ of all
contributing atoms according to \eq{eq:kick}, or to keep the all-atom velocity
constant $\vvec \rightarrow \vvec$. Generalized Newtonian event-chain Monte 
Carlo offers two schemes to treat events. The \emph{Newtonian-general} scheme 
applies to general mass matrices $\mathbf{M}$ and factors that depend on an 
arbitrary number of atoms. The \emph{Newtonian-pair} scheme only applies to 
distance-dependent pair potentials and identity mass matrix 
$\mathbf{M}=\mathbf{I}$.

In the Newtonian-general scheme, the probability for each possible choice 
$(k,\vvec')$ or $(k, \vvec)$ of the lifting variables is given by $\max[0, 
-\grad_{\xvec_k} U_{M_\text{ev}} \cdot
\vvec_k'] / C$ or  $\max[0, -\grad_{\xvec_k} U_{M_\text{ev}} \cdot \vvec_k] /
C$, respectively. Here, $C$ is a common normalization factor. If the force kick
is not applied and $\vvec$ stays constant, the active atom always changes
because the factor event rate was positive at the time of the event [see
\eqtwo{eq:event}{eq:Poisson}]. Likewise, if the event-factor potential only
depends on a single atom  (as for external potentials), the force kick is
always applied. 

In the Newtonian-pair scheme for distance-dependent pair potentials between two 
atoms $a$ and
$b$, $U_{M_\text{ev}} = U_{M_\text{ev}}(|\xvec_a - \xvec_b|)$ and an identity
mass matrix $\mathbf{M}=\mathbf{I}$, we alternatively exploit the inherent
translational invariance to treat the event deterministically. We always apply
the force kick in \eq{eq:kickdiagonal} to modify both $\vvec_a$ and $\vvec_b$,
and always change the active atom within the event factor. In this case,
\eq{eq:kickdiagonal} is equivalent to an elastic Newtonian collision between two
particles of equal mass, and this alternative scheme recovers the original
formulation of Newtonian event-chain Monte Carlo for the hard-sphere
model~(\emph{29}).

Samples of all-atom configurations $\xvec$ that can be used to compute
observables are taken at periodic time intervals~(\emph{29}).
Furthermore, Newtonian
event-chain Monte Carlo requires resamplings of the all-atom velocity $\vvec$
and the active atom $i$ in periodic time intervals $\tau_\text{chain}$ to become
irreducible~(\emph{31}). We sample $\vvec$ and $i$ from their
respective stationary distributions $\mu_\mathcal{V}(\vvec)$ and
$\mu_\mathcal{N}(i)$. For $\vvec$, this is equivalent to sampling a multivariate
Gaussian distribution with mean $\boldsymbol{\mu} = \mathbf{0}$ and covariance
matrix $\boldsymbol{\Sigma} = \mathbf{M}^{-1}$, followed by corrections of the
total velocity $\mathbf{1}^T \cdot \vvec \rightarrow 0$ and $\vvec^T \cdot
\mathbf{M} \cdot \vvec \rightarrow 2 E_\text{kin}$.
The activity label $i$ is sampled uniformly from the
the set $\mathcal{N} = \{1 \ldots N_a\}$.

\subparagraph*{Proof of Correctness of Generalized Newtonian Event-Chain Monte 
Carlo} 
\mbox{} \\
In order to prove the correctness of the Newtonian event-chain Monte
Carlo algorithm for general smooth interactions, we must show that the global
balance condition is satisfied. The total probability flow into any lifted
configuration $(\xvec, \vvec, i)\in \widehat{\Omega}$ consists of a physical
flow $\mathcal{F}^\text{phys}(\xvec, \vvec, i)$ and a lifting flow
$\mathcal{F}^\text{lift}(\xvec, \vvec, i)$ and must equal its statistical
weight $\hat{\pi}(\xvec, \vvec, i)$:
\begin{equation}
	\label{eq:global}
	\hat{\pi}(\xvec, \vvec, i) = \mathcal{F}^\text{phys}(\xvec, \vvec, i) + 
	\mathcal{F}^\text{lift}(\xvec, \vvec, i).
\end{equation}

The physical flow into $(\xvec, \vvec, i)$ stems from the continuous movement 
of the active atom $i$ that was not interrupted by an event with the event rate 
given in \eq{eq:event}. With $\xvec' = (\xvec_1, \ldots, \xvec_i - \vvec_i 
\mathrm{d}t, \ldots, \xvec_{N_a})$, we get
\begin{equation}
	\label{eq:phys}
	\begin{split}
	\mathcal{F}^\text{phys}(\xvec, \vvec, i) &= \hat{\pi}(\xvec', \vvec, i) 
	\left\{ 1 - \beta 
	\sum_M \max\left[ 0, \grad_{\xvec'_i} U_M \cdot \vvec_i \right]\right\} \\
	&= \hat{\pi}(\xvec, \vvec, i) \left\{ 1 - \beta \sum_M \max\left[0, 
	-\grad_{\xvec_i} U_M \cdot \vvec_i\right] \right\} \\
	&\eqqcolon \hat{\pi}(\xvec, \vvec, i) + \sum_M \mathcal{F}^\text{phys}_M 
	(\xvec, \vvec, i), 
	\end{split}
\end{equation}
where the second line uses the detailed-balance property of the factorized 
Metropolis filter that yields the event rates in \eq{eq:event} and allows to 
treat all factors as statistically independent~(\emph{20}).
The lifting flow into $(\xvec, \vvec, i)$ stems from interrupted motions of 
lifted configurations $(\xvec, \vvec', j)$ with different lifting variables 
$\vvec'$ and $j$ but the same configuration $\xvec$:
\begin{equation}
	\label{eq:lift}
	\begin{split}
	\mathcal{F}^\text{lift}(\xvec, \vvec, i) &= \beta\sum_M\sum_{j\in 
	\mathcal{N}} \int_{\mathcal{V}^{3N_a}(E_\text{kin})} 
	\mathrm{d}^{3N_a}\vvec'\, \hat{\pi}(\xvec, \vvec', j) \max\left[0, 
	\grad_{\xvec_j}U_M\cdot \vvec'_j\right] p^M_{(\vvec',j),(\vvec, i)} \\
	&\eqqcolon \sum_{M} \mathcal{F}^\text{lift}_M(\xvec, \vvec, i).
	\end{split}
\end{equation}
Here, $p_{(\vvec',j), (\vvec, i)}$ is the probability to change the lifting
variables $(\vvec', j)$ to $(\vvec,i)$. For every factor $M$, the lifting flow
$\mathcal{F}^\text{lift}_M(\xvec, \vvec, i)$ in \eq{eq:lift} exactly cancels
the physical flow $\mathcal{F}^\text{phys}_M(\xvec, \vvec, i)$ in \eq{eq:phys}
so that the global-balance condition in \eq{eq:global} is satisfied. A trivial 
(inefficient) solution would be to simply invert the velocity of the
active atom $i$ in an event while keeping it active: $p^M_{(\vvec',j),(\vvec,i)}
= \delta_{ij} \, \delta^{(3)}(\vvec_i - \vvec'_i)$.

We presented two schemes for the update of the lifting variables in an event
of generalized Newtonian event-chain Monte Carlo. The first stochastic 
Newtonian-general scheme 
applies for a
general number of atoms on the event factor. The second deterministic 
Newtonian-pair scheme 
exploits the translational invariance of distance-dependent pair potentials. 
For the Newtonian-general scheme, we get
\begin{equation}
	\label{eq:stochastic}
	\begin{split}
	p^M_{(\vvec', j), (\vvec, i)} = &\frac{\max\left[ 0, -\grad_{\xvec_i} U_M 
	\cdot \vvec_i \right]}{C} \bigg[ \\
	&\quad\delta^{(3N_a)}\left( \vvec - \vvec' + 2 
	\, \frac{(\vvec')^T \cdot \grad_{\xvec} 
		U_{M}}{(\grad_{\xvec}U_{M})^T \cdot 
		\mathbf{M}^{-1} \cdot \grad_{\xvec}U_{M}} \, 
	\mathbf{M}^{-1}\cdot\grad_{\xvec} 
	U_{M}
	\right) \\
	&\quad+ \delta^{(3N_a)}\left(\vvec - \vvec'\right) \bigg] \\
	= &\frac{\max\left[ 0, -\grad_{\xvec_i} U_M 
	 	\cdot \vvec_i \right]}{C} \bigg[ \\
	 &\quad\delta^{(3N_a)}\left( \vvec' - \vvec + 2 
	 \, \frac{\vvec^T \cdot \grad_{\xvec} 
	 	U_{M}}{(\grad_{\xvec}U_{M})^T \cdot 
	 	\mathbf{M}^{-1} \cdot \grad_{\xvec}U_{M}} \, 
	 \mathbf{M}^{-1}\cdot\grad_{\xvec} 
	 U_{M}
	 \right) \\
	 &\quad+ \delta^{(3N_a)}\left(\vvec' - \vvec\right) \bigg],
	\end{split}
\end{equation}
where the first term applies the force kick, while the second does not. 
From the necessary condition
\begin{equation}
	1 = \sum_{k\in\mathcal{N}} \int_{\mathcal{V}^{3N_a}(E_\text{kin})} 
	\mathrm{d}^{3N_a}\vvec\, p^M_{(\vvec',j),(\vvec,k)},
\end{equation}
it follows that
\begin{equation}
	\label{eq:c}
	\begin{split}
	C &= \sum_{k\in\mathcal{N}}\bigg\{\max\left[0, -\grad_{\xvec_k} U_M \cdot 
	\left( \vvec'_k - 2\frac{(\vvec')^T \cdot \grad_\xvec U_M}{(\grad_\xvec 
	U_M)^T \cdot \mathbf{M}^{-1}\cdot \grad_\xvec U_M} 
	\left(\mathbf{M}^{-1}\cdot 
	\grad_\xvec U_M\right)_k 
	\right)\right] \\
	&\quad\quad\quad\,\,\,\,+ \max\left[0, 
	-\grad_{\xvec_k} U_M \cdot 
	\vvec_k'\right] \bigg\}\\
	&= \sum_{k\in\mathcal{N}}\bigg\{\max\left[0, \grad_{\xvec_k} U_M \cdot 
	\left( \vvec'_k - 2\frac{(\vvec')^T \cdot \grad_\xvec U_M}{(\grad_\xvec 
	U_M)^T \cdot \mathbf{M}^{-1}\cdot \grad_\xvec U_M} 
	\left(\mathbf{M}^{-1}\cdot 
	\grad_\xvec U_M\right)_k 
	\right)\right] \\
	&\quad\quad\quad\,\,\,\,+ \max\left[0, 
	\grad_{\xvec_k} U_M \cdot 
	\vvec_k'\right] \bigg\}.
	\end{split}
\end{equation}
Here, the second rewriting follows from expressing the trivial equality
\begin{equation}
	\vvec^T\cdot \grad_\xvec U_M + \left(\vvec - 2 \frac{\vvec^T \cdot 
	\grad_\xvec U_M}{(\grad_\xvec U_M)^T \cdot \mathbf{M}^{-1} \cdot 
	\grad_\xvec 
	U_M} \mathbf{M}^{-1} \cdot \grad_\xvec U_M \right)^T \cdot \grad_\xvec U_M 
	= 0
\end{equation}
as a sum over $k\in \mathcal{N}$ and from grouping the positive and negative
terms separately.
The normalization factor $C$ in $p^M_{(\vvec', j), (\vvec, i)}$ is the same for
every possible next active atom $i$. Because of the $\delta^{(3 
N_a)}$-functions in \eq{eq:stochastic}, the normalization factor $C$ can also 
be written in terms of the velocity $\vvec$ after the force kick by replacing 
$\vvec'\rightarrow\vvec$ in \eq{eq:c}. This yields
\begin{equation}
	\label{eq:liftphys}
	\begin{split}
	\mathcal{F}^\text{lift}_M(\xvec, \vvec, i) &= \beta \, 
	\hat{\pi}(\xvec, \vvec, i) \max\left[0, -\grad_{\xvec_i} U_M \cdot 
	\vvec_i\right] \frac{1}{C} \sum_{j\in \mathcal{N}}  \bigg\{ \\
	&\hphantom{=}\quad\max\left[0, \grad_{\xvec_j} U_M \cdot \left( 
	\vvec_j - 2 \frac{\vvec^T \cdot \grad_\xvec U_M}{(\grad_\xvec U_M)^T \cdot 
	\mathbf{M}^{-1} \cdot \grad_\xvec U_M} \left(\mathbf{M}^{-1} \cdot 
	\grad_\xvec U_M\right)_j\right)\right] \\
	&\hphantom{ = }\quad + \max\left[ 0, \grad_{\xvec_j} U_M \cdot \vvec_j 
	\right] \bigg\} \\
	&= \beta \, \hat{\pi}(\xvec, \vvec, i) \max\left[0, -\grad_{\xvec_i} U_M 
	\cdot \vvec_i\right] \\
	&= - \mathcal{F}^\text{phys}_M(\xvec, \vvec, i),
	\end{split}
\end{equation}
where we used that the velocity $\vvec$ and activity label $i$ are
uniformly distributed over their sample spaces $\mathcal{V}^{3 
N_a}(E_\text{kin})$ and 
$\mathcal{N}$, respectively. \Eq{eq:liftphys} concludes the proof that the 
presented Newtonian-general scheme for the update of the lifting variables in 
an 
event by a general factor potential 
satisfies the global-balance condition of \eq{eq:global}. The
reflective~(\emph{23}) and 
forward~(\emph{30}) event-chain Monte Carlo variants can likewise be
generalized to
factor potentials depending on an arbitrary number of atoms.

For the Newtonian-pair scheme for distance-dependent pair potentials and 
identity mass matrix $\mathbf{M}=\mathbf{I}$, we 
get
\begin{equation}
	\begin{split}
	p^M_{(\vvec', j), (\vvec, i)} &=
	\begin{cases}
		\delta^{(3N_a)}\left( \vvec - \vvec' + 2 
		\, \frac{(\vvec')^T \cdot \grad_{\xvec} 
			U_{M}}{(\grad_{\xvec}U_{M})^T \cdot \grad_{\xvec}U_{M}} \, 
		\grad_{\xvec} 
		U_{M}
		\right) & \text{if $i,j\in M$ and $i \neq j$,} \\
		0 & \text{otherwise,}
	\end{cases} \\
	&= \begin{cases}
		\delta^{(3N_a)}\left( \vvec' - \vvec + 2 
		\, \frac{\vvec^T \cdot \grad_{\xvec} 
			U_{M}}{(\grad_{\xvec}U_{M})^T \cdot 
			\grad_{\xvec}U_{M}} \, 
		\grad_{\xvec} 
		U_{M}
		\right) & \text{if $i,j\in M$ and $i \neq j$,} \\
		0 & \text{otherwise.}
	\end{cases}	
	\end{split}
\end{equation}
With $i,j\in M$ and $i\neq j$, this yields
\begin{equation}
	\begin{split}
	\mathcal{F}^\text{lift}_M(\xvec,\vvec,i) &= \beta \, 
	\hat{\pi}(\xvec,\vvec,i) 
	\max\left[0, \grad_{\xvec_j} U_M \cdot \left( \vvec_j - 2 \frac{\vvec^T 
	\cdot \grad_\xvec U_M}{(\grad_\xvec U_M)^T \cdot \grad_{\xvec} U_M} 
	\grad_{\xvec_j} U_M 
	\right)\right] \\ 
	&= \beta \, \hat{\pi}(\xvec,\vvec,i) \max\left[0, \grad_{\xvec_j} U_M \cdot 
	\vvec_j - 2 \frac{\vvec_i^T \cdot \grad_{\xvec_i} U_M + \vvec_j^T\cdot 
	\grad_{\xvec_j}U_M}{\left|\grad_{\xvec_i} U_M\right|^2 + 
	\left|\grad_{\xvec_j}U_M\right|^2} \left| \grad_{\xvec_j} U_M\right|^2 
	\right] \\
	&= \beta \, \hat{\pi}(\xvec,\vvec,i) \max\left[ 0, -\grad_{\xvec_i} U_M 
	\cdot \vvec_i \right] \\
	&= -\mathcal{F}^\text{phys}_M (\xvec, \vvec, i).
	\end{split}
\end{equation}
Here, we again used the uniform distributions of $\vvec$ and $i$ and the
translational invariance $\grad_{\xvec_i}U_M = -\grad_{\xvec_j} U_M$.
Thus, also the Newtonian-pair scheme satisfies the
global-balance condition of \eq{eq:global}. We note that it always applies the 
force kick.

\subparagraph*{Fibonacci Vectors} \mbox{} \\
The cell-veto algorithm was previously only used for the straight variant
of event-chain Monte Carlo~(\emph{34, 28, 32}).
Its finite
set of possible velocities $\mathcal{D}$ of the single active atom $a$ 
allows to set up a Walker table for every velocity $\vvec_a\in\mathcal{D}$. For 
instance, for factor pair potentials between the active atom $a$ and another 
atom $b$ at positions $\xvec_a$ and $\xvec_b$, $U_M(\xvec) = U_M(\xvec_a, 
\xvec_b)$, every cell bound 
$q_M^\text{cell}(\mathcal{C}_a, \mathcal{C}_b, \vvec_a)$ for the pair of cells 
$\mathcal{C}_a$ and $\mathcal{C}_b$ in the Walker table may be  written as
\begin{equation}
\label{eq:cell}
	q_M^\text{cell}(\mathcal{C}_a, \mathcal{C}_b, \vvec_a) = \max_{\xvec_a \in 
	\mathcal{C}_a, \xvec_b \in \mathcal{C}_b} \beta \max\left[0, 
	\grad_{\xvec_a} U_M 
	(\xvec_a, \xvec_b) \cdot 
	\vvec_a\right].
\end{equation}
In principle, a Walker table must be precomputed for every velocity
$\vvec_a\in\mathcal{D}$ and for every possible cell $\mathcal{C}_a$ of the
active atom. Symmetries, such as a translational invariance of the factor
potential, may heavily reduce the necessary number of Walker
tables~(\emph{28}). During the straight event-chain Monte Carlo
simulation, the relevant Walker table is determined by the cell of the currently
active atom and its velocity. The generalization of \eq{eq:cell} to more complex
factors is straightforward. For example, in order to treat the Coulomb
interaction between two water molecules in the SPC/Fw water model, the position
and orientation of the water molecule containing the active atom may be varied
in $\mathcal{C}_a$, while the position and orientation of the other molecule is
varied in $\mathcal{C}_b$. From a practical point of view, the cell bounds
$q_M^\text{cell}(\mathcal{C}_a, \mathcal{C}_b)$ are usually not computed exactly
but rather approximated, e.g., by considering a finite set of positions
$\xvec_a\in\mathcal{C}_a$ and $\xvec_{b}\in\mathcal{C}_b$, and including a
corrective multiplicative prefactor~(\emph{32}). As long as the
approximated cell bound satisfies $\tilde{q}^\text{cell}_M (\mathcal{C}_a,
\mathcal{C}_b, \vvec_a) \geq q^\text{cell}_M(\mathcal{C}_a,\mathcal{C}_b,
\vvec_a)$, a Poisson thinning procedure~(\emph{36}) corrects
\emph{any} overestimate. The quality of $\tilde{q}^\text{cell}_M
(\mathcal{C}_a, \mathcal{C}_b, \vvec_a)$, however, does influence
the performance because higher cell bounds in the cell-veto algorithm yield 
more events per unit distance that have to be confirmed by 
computing the actual event rate. 

The inherent discretization of continuous space in \eq{eq:cell} can be 
translated to a continuous velocity space, as it
appears, e.g., in Newtonian event-chain Monte Carlo. Consider a finite number 
of unit vectors $\hat{\vvec}_d$ on the two-dimensional unit sphere, and let 
$\mathcal{V}_d$ be the
associated Voronoi cells under some distance function. We can then 
formally compute the cell bounds for every Voronoi cell as
\begin{equation}
	q_M^\text{cell}(\mathcal{C}_a, \mathcal{C}_b, \mathcal{V}_a) = 
	\max_{\xvec_a\in\mathcal{C}_a, \xvec_b 
	\in\mathcal{C}_b,\vvec_a\in\mathcal{V}_a} \beta \max\left[0, 
	\grad_{\xvec_a} U_M (\xvec_a, \xvec_b) \cdot \vvec_a\right].
\end{equation}
The results are used to precompute Walker tables for every Voronoi cell
$\mathcal{V}_a$ and for every cell $\mathcal{C}_a$, where symmetries may again 
heavily reduce the number of actually necessary tables. During the Newtonian 
event-chain Monte Carlo simulation, the relevant Walker table is determined by 
finding the cell of the currently active atom and the Voronoi cell of its 
normalized velocity $\vvec_a / |\vvec_a|$. We can then correct all cell bounds 
in the relevant Walker table to the actual speed of the active atom 
by multiplying them with $|\vvec_a|$. 

In this paper, as proof of concept, we map a generalized Fibonacci lattice onto 
a two-dimensional unit sphere by the Lambert cylindrical equal-area
projection to generate $D$ unit vectors $\hat{\vvec}_d^\text{fib}$:
\begin{equation}
\label{eq:fib1}
	\hat{\vvec}_d^\text{fib} = \left(\cos\phi_d \sin\theta_d, 
	\sin\phi_d\sin\theta_d, \cos\theta_d \right)^T,
\end{equation}
where $0\leq d < D$ and
\begin{equation}
\label{eq:fib2}
	\begin{split}
	\phi_d &= \frac{2\pi d}{\varphi},\\
	\theta_d &= \arccos \left(1 - \frac{2 (d + \varepsilon)}{D - 1 + 
	2\varepsilon}\right),
	\end{split}
\end{equation}
with the golden ratio $\varphi = (1 + 
\sqrt{5})/2$~(\emph{60, 61, 62}). The empirical 
choice of the parameter $\varepsilon=\num{0.36}$ optimizes the average
nearest-neighbor distance of the Fibonacci vectors $\hat{\vvec}_d$. As a
distance function to construct the Voronoi cells $\mathcal{V}_d$, we use 
the quick-to-evaluate cosine distance.
\Eqtwo{eq:fib1}{eq:fib2}
efficiently generate an arbitrary number of Fibonacci 
vectors with a nearly uniform distribution. However, there is no efficient 
inverse mapping from a general velocity $\vvec_a$ 
to the closest Fibonacci vector $\hat{\vvec}_d$. This is not a problem in this 
paper because we choose $D=10$ small. Then, $\hat{\vvec}_d$ can be found by
brute force. If considerably larger values of $D$ become necessary,
other point configurations on the two-dimensional unit sphere may 
be considered~(\emph{62, 63}). For the uniform SPC/Fw water 
systems in this paper, the Walker tables do not strongly depend on the 
respective Fibonacci vector. The proper discretization of velocity space will, 
however, impact  the performance of Newtonian event-chain
Monte Carlo in nonuniform systems.

\renewcommand\thefigure{S\arabic{figure}}    

\begin{figure}[b]
	\centering
	\includegraphics[width=10.89cm]{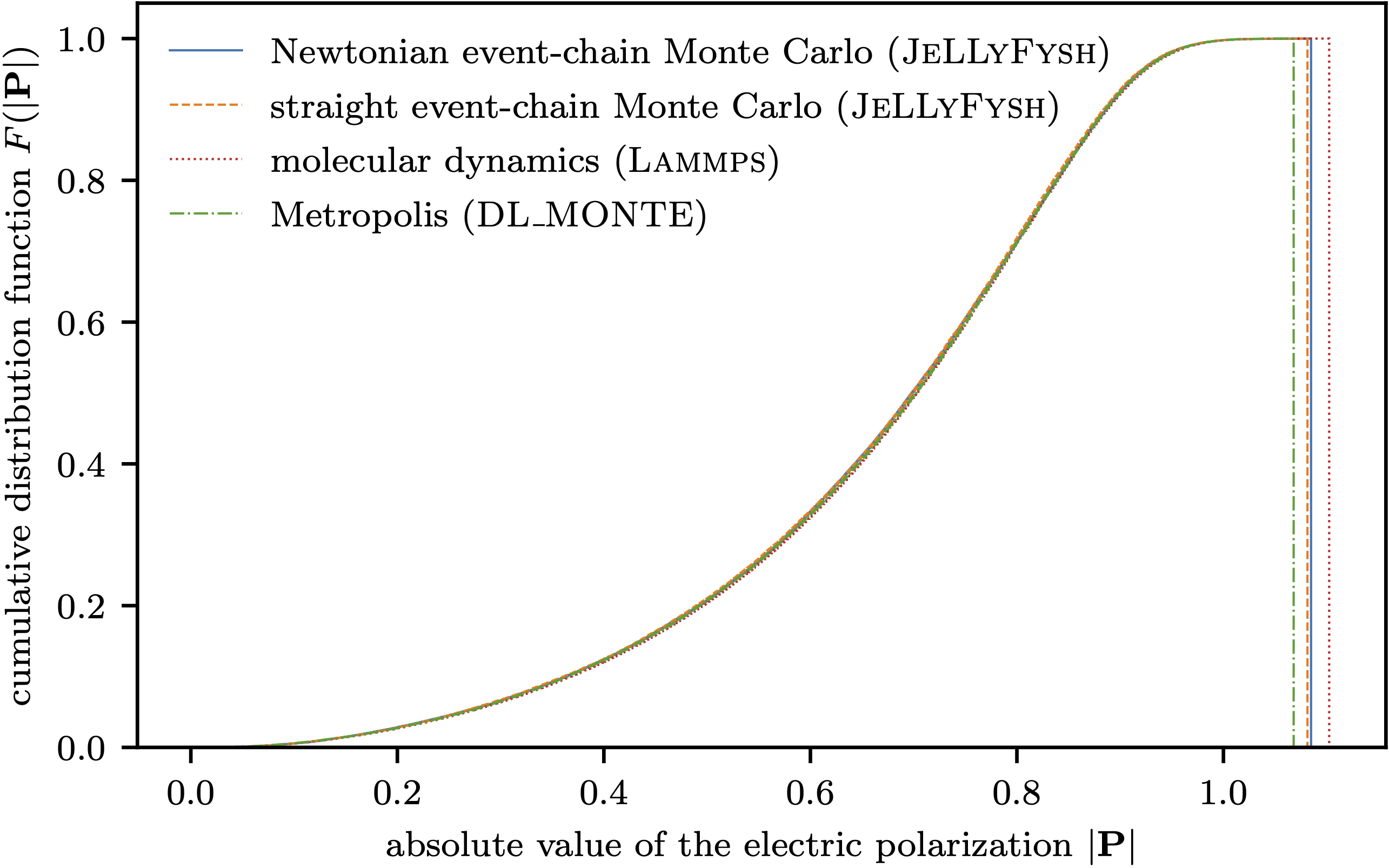}
	\caption{\textbf{Validation for the SPC/Fw water model.} 
	Cumulative distribution function of the absolute value of the 
	electric polarization of $N=2$ SPC/Fw water molecules in a cubic box of 
	side length $L=\SI{20}{\angstrom}$. The simulation protocols of the 
	different sampling methods are described in the \nameref{sec:materials}.}
	\label{fig:Validation}
\end{figure}


